\documentclass[final,5p,times,twocolumn]{elsarticle}
\usepackage{amsmath}
\usepackage{amsfonts}
\usepackage{amssymb}
\usepackage{graphicx}
\usepackage{dcolumn}
\usepackage{color}
\journal{Arxiv}

\begin{document}

\begin{frontmatter}

\title{Analytic solution for the buckling of a short cylindrical
shell.\\ Why old tires are still being preferred as dock bumpers in
harbours}

\author[utal]{Miguel Lagos}
\ead{mlagos@utalca.cl}
\address[utal]{Facultad de Ingenier\'\i a, Universidad de Talca,
Campus Los Niches, Camino a los Niches Km 1, Curic\'o, Chile}

\begin{abstract}
The profile of a cylindrical shell in the post--buckling regime of
deformation is calculated exactly, for any load and axial deformation,
and is shown to be a Jacobi elliptic sine function. Cylindrical
symmetry is assumed from the beginning. Results shed light on why old
tires are so effective as dock and tugboat bumpers from long ago to
nowadays, proving to be safe for berthing even the largest
freighters. As no technical analysis of this capacity has been
published yet, it seems opportune to apply the mathematical solution
to explain this long lasting enigma. The reaction force and stored
energy of an axially compressed cylindrical shell, as a tire tread is,
exhibit ideal behaviour for a safety shock energy absorber. Most
energy absorption takes place after buckling, which bends the
cylindrical mantle and increases its circumference, inducing strong
tensile forces that contribute to stabilize the buckled structure.
\end{abstract}


\end{frontmatter}

\section{Introduction}
\label{introd}

There is a great deal of work done on the critical conditions for the
buckling of narrow or thin walled solids of a variety of shapes, and
on the subsequent evolution of the collapsing element \cite{Roarks}.
Buckling is generally the controlling failure mode of thin shell
structures and long narrow structural elements, hence the study of the
elastic instabilities of shells is a most important subject of
materials mechanics and engineering design. There is an extensive list
of books, review articles and conference proceedings providing a
wealth of information on the strength, mechanical stability and
buckling behavior of thin shell structures \cite{Teng}. However,
either explicitly or not, most of these studies have associated the
idea of buckling as a catastrophic collapse of the critically strained
thin walled or narrow element. Also, the mathematical treatments are
often intended to be as general as possible
\cite{Hunt,Paschero,Pinna,Simitnes,Wullschleger}, which precludes
the obtention of closed--form practical formulas of use in specific
practical situations, and able of being tabulated in manuals
\cite{Roarks}.

This communication deals with a mechanical model for a very particular
technical application in which buckling is not an undesired effect,
but a highly beneficial one. The use of old tires as dock bumpers has
been an old tradition in most harbours. Although they are now being
replaced by specially fabricated marine fenders, ship captains
appreciate the behaviour of old tires in docking because, for
instance, they provide a characteristic smooth absorption of the ship
kinetic energy with total absence of spring back at the end of the
operation. The question of why old tires are still being preferred as
dock protections, even for the docking of large cargo vessels, is an
enigma whose precise solution is the subject of what follows.

\begin{figure}[h!]
\begin{center}
\includegraphics[width=7cm]{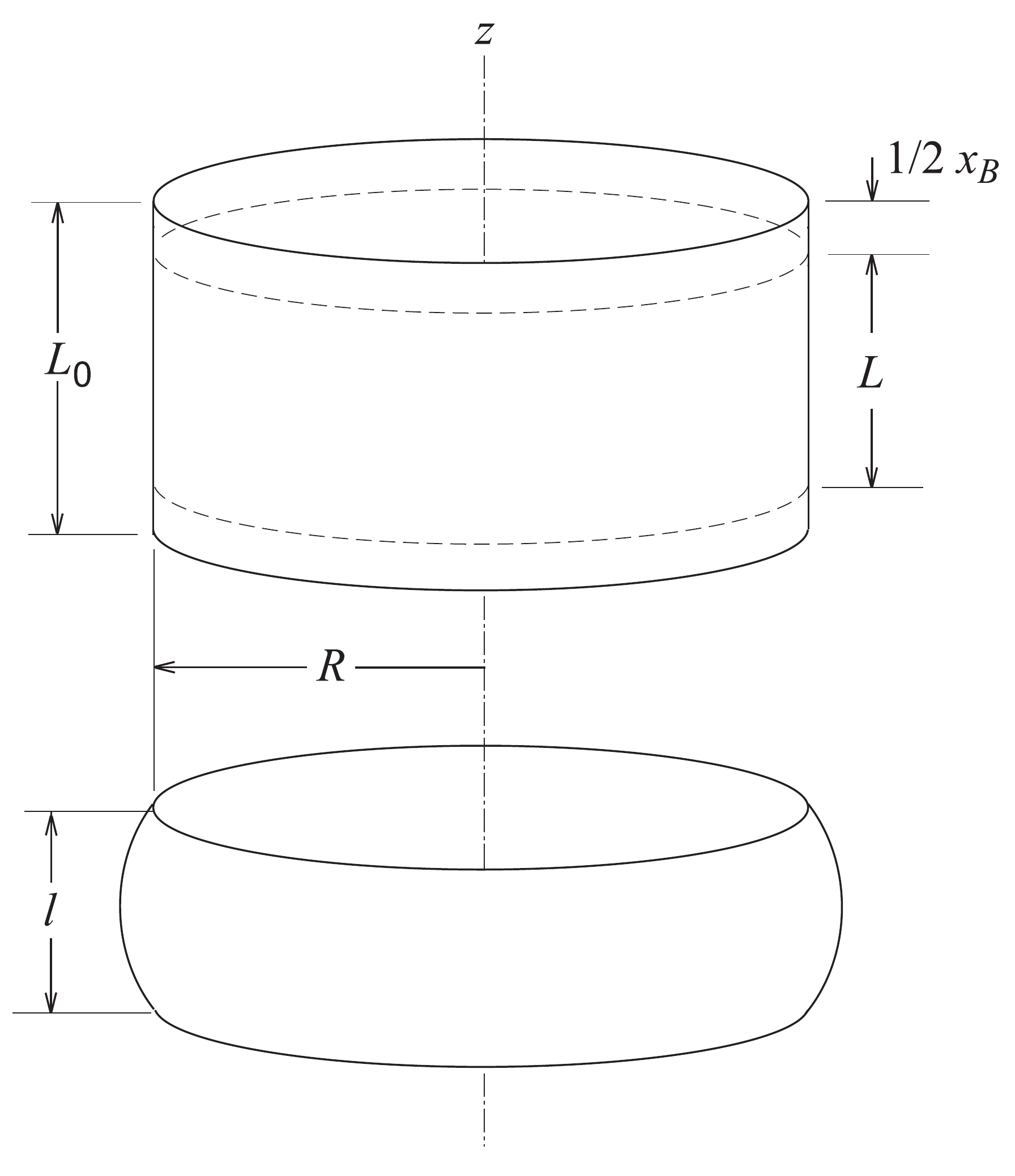}
\caption{\label{Fig1} Model for the tread of a pneumatic tire as a
cylindrical shell. (a) Unstrained condition (solid lines) and critical
axial strain for buckling (discontinuous lines). (b) Buckled strain
regime by the action of compressive axial forces homogeneously
distributed on the shell edges.}
\end{center}
\end{figure}

The reaction force and energy absorption effect of an axially
compressed tire is attributed entirely to the elastic deformation of
the tread, which is modeled here as a cylindrical shell whose radius
$R$ is greater than its axial length $L_0$, both magnitudes measured
in the unstrained condition, as shown schematically in
Fig.~\ref{Fig1}. In opposition to small tires, whose treads are often
initially slightly barreled, big ones are essentially cylindrical when
unloaded.  Though small, the shell thickness $e$ is assumed large
enough to preclude the formation of patterns of complex geometry upon
deformation. The applied opposed forces of strength $P$ are parallel
to the symmetry $z$--axis and are uniformly distributed along the
edges of the cylindrical mantle. It is also assumed that the edges of
the elastic cylinder are not deformed because they rest on rigid plane
surfaces and the friction forces are strong enough to prevent sliding.
Hooke's law is assumed to hold in a first approach. Although rubber
elastic properties exhibit deviations from linear response, the main
conclusions will not be much affected by this.

Instead of dealing with the model in the most general way, facing all
the complex deformation modes a cylindrical shell may display
\cite{Hunt,Paschero,Pinna,Simitnes,Wullschleger}, the mathematical
approach is kept as simple as possible by allowing only barrel shaped
deformations. The equilibrium equations of the model incorporate from
the very beginning just the deformation modes one can observe in an
axially compressed tire, and discards any other because the main
objective is to understand the ability of tires as shock energy
absorbers.

The mechanical analysis shows that, in a short first stage, the
cylindrical shell undergoes a uniform compressive strain along the
$z$--axis, conserving strictly its cylindrical shape and opposing a
reaction force proportional to the strain up to a maximal load $P_B$.
The maximum strain reached in this first deformation regime is
indicated in Fig.~\ref{Fig1}(a) by a discontinuous line. Compression
beyond this limit makes the shell to start buckling, and the
cylindrical mantle progressively acquires a barrel shape, as in
Fig.~\ref{Fig1}(b). In this second regime the reaction force
diminishes monotonically with strain, but conserving always a
significant strength. The development of the barrel profile involves
stretching of the tread circumference, which demands strong tensile
forces that contribute to stabilize the buckled structure. As a shock
energy absorber, the system exhibits the advantages of having a
maximum reaction force, which can be designed to warrant no damage to
the colliding bodies, together with a large energy absorption
capability, particularly in the buckled strain regime.

\section{Buckled strain regime of the cylindrical shell of finite
thickness}
\label{buckling}
\subsection{Equilibrium of the internal forces}
\label{forces}

\begin{figure}[h!]
\begin{center}
\includegraphics[width=7cm]{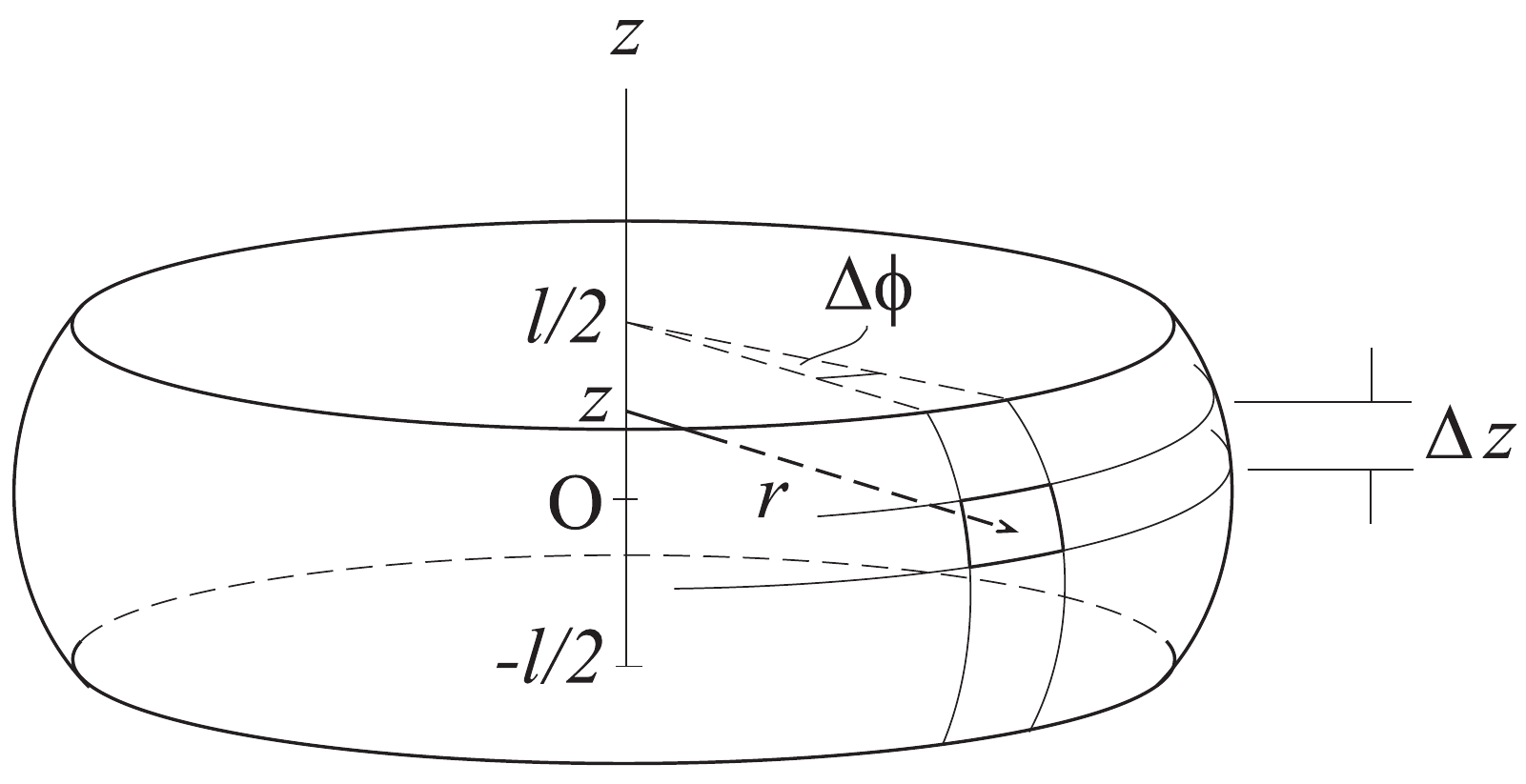}
\caption{\label{Fig2} The strained cylindrical shell, showing
schematically an elementary sector defined by the coordinates $r$,
$\phi$ and $z$ ($\phi$ is implicit), and by the variations
$\Delta\phi$ and $\Delta z$.}
\end{center}
\end{figure}

Fig.~\ref{Fig2} shows the cylindrical shell in the strain regime in
which it adopts a barrel shaped profile by effect of the axial load
$P$ (equivalent to the reaction force at equilibrium, or
quasi--equilibrium). The cylindrical coordinate system has its
$z$--axis along the main symmetry axis and the origin $O$ at the
center of the deformed shell, which extends from $z=-l/2$ to
$z=l/2$. Distance $r$ is the radius of the shell for a given $z$, so
that function $r(z)-R$ determines the profile of the strained
shell. The azimuthal coordinate $\phi$ is the usual one in cylindrical
coordinates, and is not shown in Fig.~\ref{Fig2} for the sake of
clarity.

\vskip 10pt

The equilibrium equations are derived from the analysis of the forces
operating on an elementary sector of the shell, schematically
represented in Fig.~\ref{Fig2}. The method is preferred over simply
writing the equations of the theory of elasticity and applying
boundary conditions because is simpler and provides better physical
insight. The advantages of the procedure are discussed in the final
subsection of this section.

\begin{figure}[h!]
\begin{center}
\includegraphics[width=3.5cm]{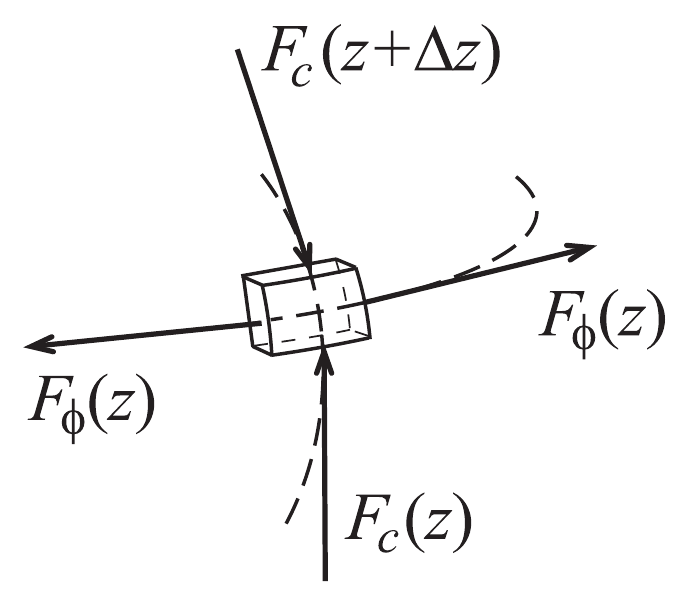}
\caption{\label{Fig3} The elementary sector of Fig.~\ref{Fig2} and the
compressive and tensile forces $F_c$ and $F_\phi$ exerted on it.}
\end{center}
\end{figure}

Fig.~\ref{Fig3} displays a diagram of the material element and the
forces applied on it. Forces in the plane $\phi =\text{constant}$,
containing the $z$--axis, are compressive forces of strength
$F_c(z+\Delta z)$ and $F_c(z)$. The forces in the plane
$z=\text{constant}$, normal to the $z$--axis, are tensile forces
because stretch the shell contour in its plane, and their strength is
denoted $F_\phi(z)$. Solutions with cylindrical symmetry were
implicitly assumed because no dependence of the forces on $\phi$ was
considered.

\begin{figure}[h!]
\begin{center}
\includegraphics[width=5.3cm]{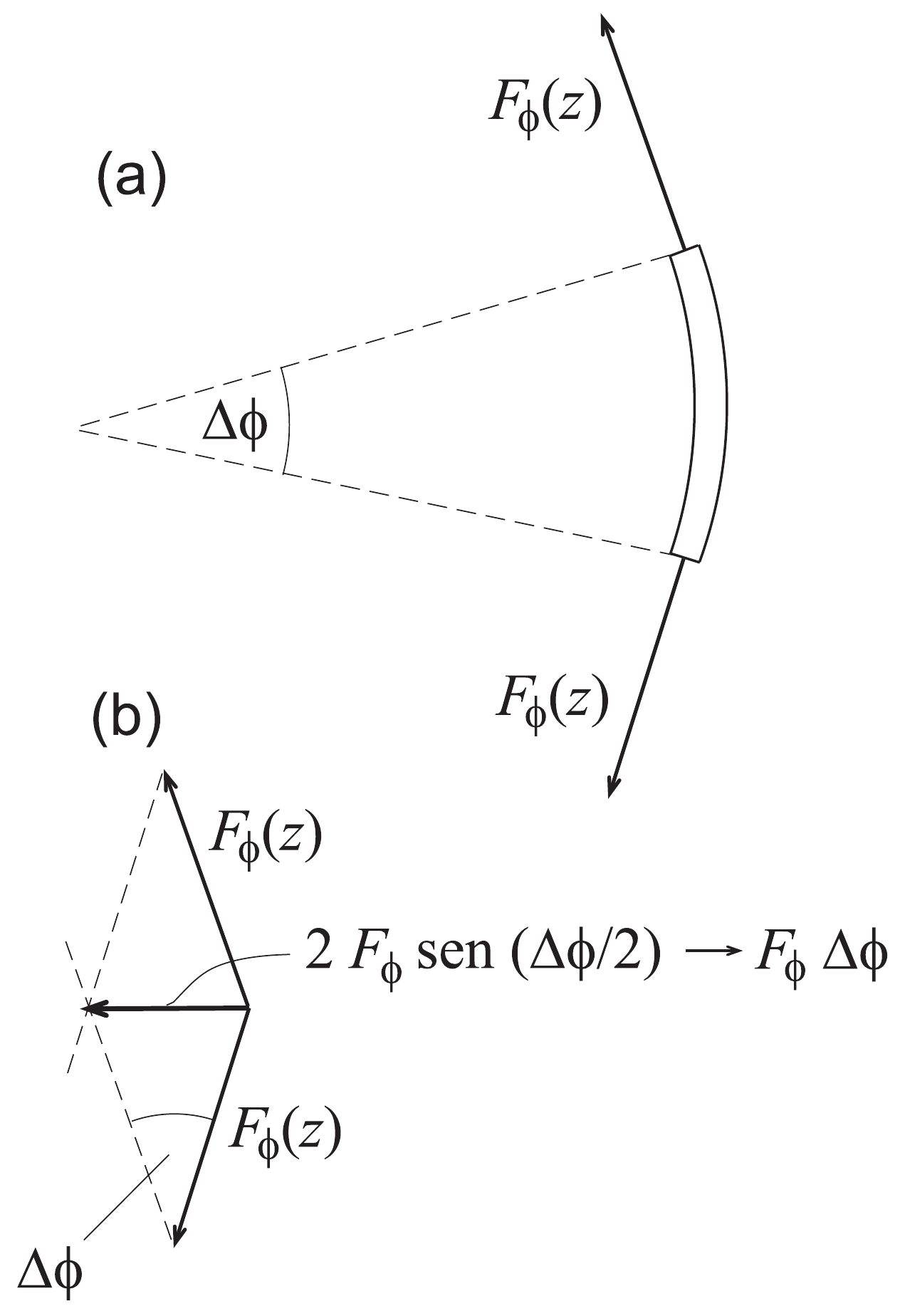}
\caption{\label{Fig4} (a) The graphic scheme of Fig.~\ref{Fig3}
projected on the plane $z=\text{constant}$. (b) Force diagram in the
plane $z=\text{constant}$.}
\end{center}
\end{figure}

Fig.~\ref{Fig4}(a) is a projection on the plane $z=\text{constant}$ of
the system of Fig.~\ref{Fig3} showing the forces operating in this
plane, and Fig.~\ref{Fig4}(b) shows the vector composition of them.
Hence, the tensile forces give a sum of strength

\begin{equation}
2F_\phi(z)\,\sin (\Delta\phi /2) 
\rightarrow F_\phi(z)\Delta\phi
\;\,\text{if}\;\,\Delta\phi\approx 0,
\label{Ec1}
\end{equation}

\noindent
contained in the plane $z=\text{constant}$ and pointing towards the
central $z$--axis. 

\begin{figure}[h!]
\begin{center}
\includegraphics[width=5.3cm]{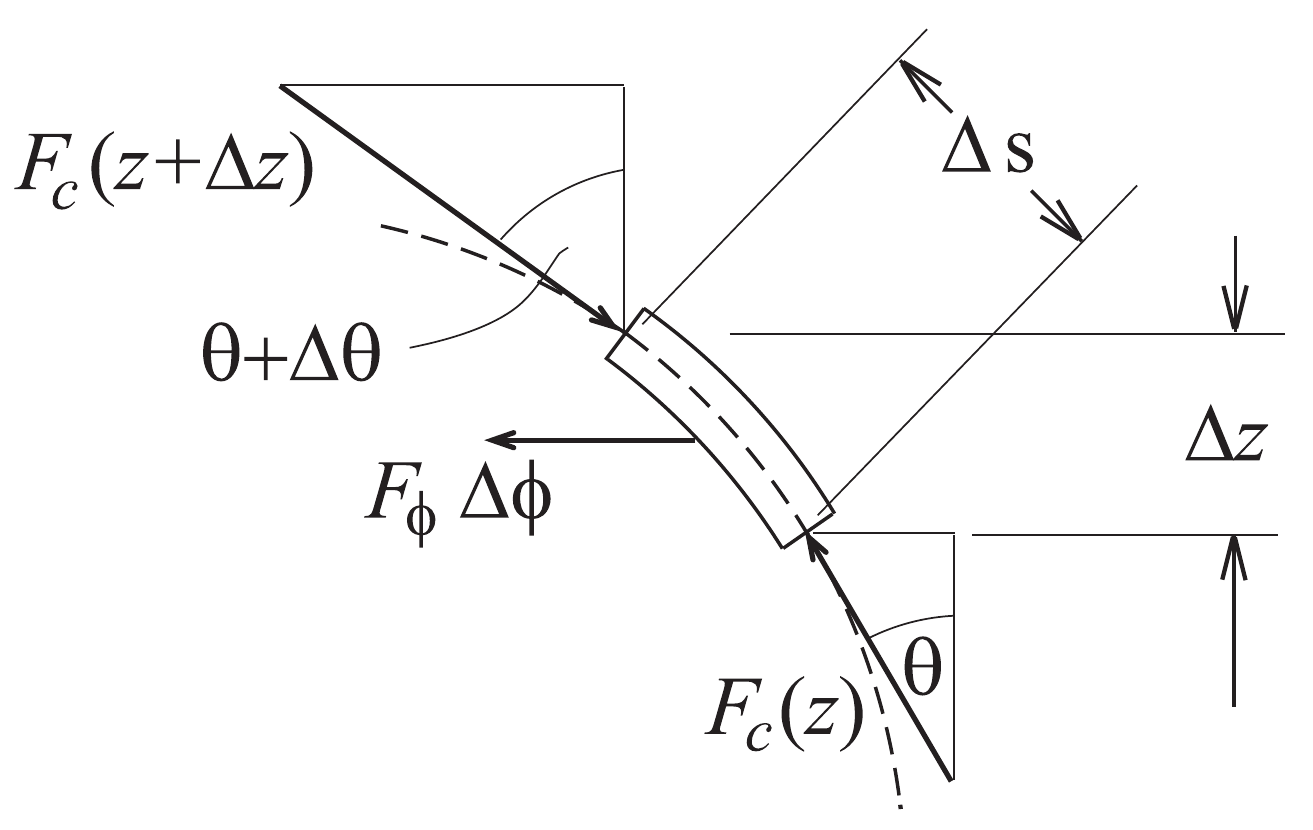}
\caption{\label{Fig5} Forces in the plane $\phi =\text{constant}$.
Compressive forces $F_c(z+\Delta z)$ and $F_c(z)$ are tangent to the
curve $r(z)-R$ representing the shell profile, and subtend angles
$\theta +\Delta\theta$ and $\theta$ with the $z$--axis, respectively.
The force $F_\phi(z)\Delta\phi$ is obtained in Fig.~\ref{Fig4} from
the tensile forces in the plane $z=\text{constant}$. The discontinuous
line represents the shell profile.}
\end{center}
\end{figure}

Forces in the plane $\phi =\text{constant}$ are shown in
Fig.~\ref{Fig5}. They are the two compresive forces of strength
$F_c(z+\Delta z)$ and $F_c(z)$ operating in the two opposite edges,
and the third one is the resultant of the tensile forces in the plane
$z=\text{constant}$, whose strength is $F_\phi(z)\Delta\phi$. The
former two forces are tangent to the curve $r(z)-R$ determining the
shell profile in the corresponding application points, and subtend
angles $\theta +\Delta\theta$ and $\theta$ with the $z$--axis,
respectively.

Equilibrium demands

\begin{equation}
-F_c(z+\Delta z)\cos (\theta +\Delta\theta)+F_c(z)\cos\theta=0,
\label{Ec2}
\end{equation}

\begin{equation}
F_c(z+\Delta z)\,\sin (\theta +\Delta\theta)
-F_c(z)\,\sin\theta=-F_\phi\Delta\phi .
\label{Ec3}
\end{equation}

\begin{figure}[h!]
\begin{center}
\includegraphics[width=3cm]{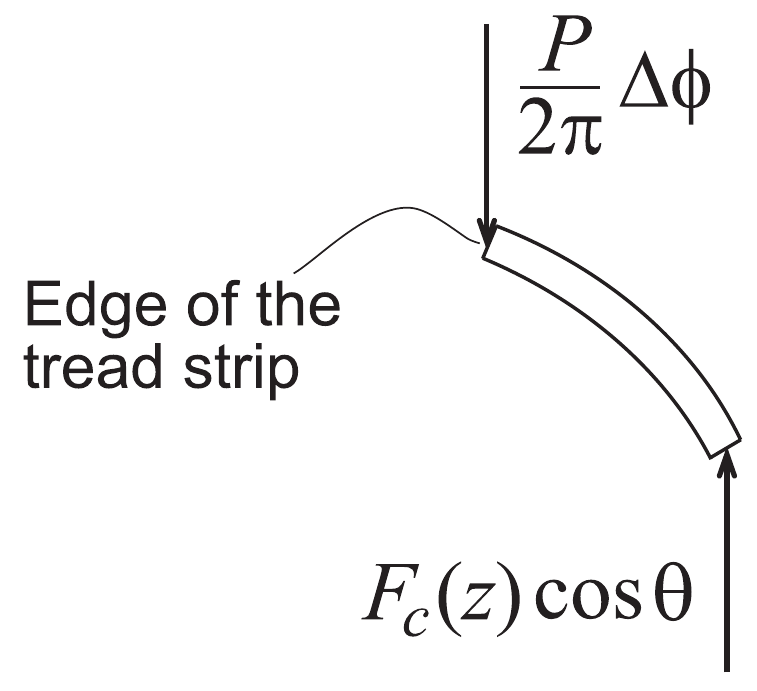}
\caption{\label{Fig6} Equilibrium of a shell fringe of finite size in
the $z$--direction. The $z$--component $F_c(z)\cos\theta$ of the force
is $P\Delta\phi /(2\pi)$ for any $z$ in $[-l/2,l/2]$.}
\end{center}
\end{figure}

\noindent
From Eq.~(\ref{Ec2}) one can infer that the $z$--component of the
forces does not depend on $z$, and hence they can be identified with
the external force $P\Delta\phi /(2\pi)$ applied on the edge of the
shell fringe defined by $\Delta\phi$. The argument is graphically
shown in Fig.~\ref{Fig6} and yields

\begin{equation}
F_c(z+\Delta z)\cos (\theta +\Delta\theta)
=F_c(z)\cos\theta
=\frac{P}{2\pi}\Delta\phi.
\label{Ec4}
\end{equation}

\subsection{Differential equation for the shell profile function
$r(z)-R$}
\label{ecuation}

Recalling Eqs.~(\ref{Ec4}) one can divide the first term in the right
hand side of Eq.~(\ref{Ec3}) by $F_c(z+\Delta z)\cos (\theta
+\Delta\theta)$, the second one by $F_c(z)\cos\theta$, and the right
hand side by $P\Delta\phi /(2\pi)$.  It gives

\begin{equation}
\tan(\theta +\Delta\theta)-\tan\theta
=-\frac{2\pi F_\phi}{P} .
\label{Ec5}
\end{equation}

\noindent
Force $F_\phi$ can be expressed as

\begin{equation}
F_\phi =\sigma_{\phi\phi} e\Delta s ,
\label{Ec6}
\end{equation}

\noindent
where $\sigma_{\phi\phi}$ is the normal stress in the azimuthal
direction, $e$ is the shell thickness and $\Delta s$ the length of the
shell element shown in Fig.~\ref{Fig5}, and whose value is

\begin{equation}
\Delta s=\sqrt{1+\left(\frac{dr}{dz}\right)^2}\,\Delta z.
\label{Ec7}
\end{equation}.

\noindent
Replacing $\tan\theta =dr/dz$ and combining Eqs.~(\ref{Ec5}),
(\ref{Ec6}) and (\ref{Ec7}) one obtains that, in the limit $\Delta
z\rightarrow 0$,

\begin{equation}
\frac{d^2r}{dz^2}
=-\frac{2\pi e\sigma_{\phi\phi}}{P}
\sqrt{1+\left(\frac{dr}{dz}\right)^2}.
\label{Ec8}
\end{equation}

On the other hand, the normal stresses $\sigma_{rr}$,
$\sigma_{\phi\phi}$ and $\sigma_{zz}$ satisfy Hooke's law

\begin{equation}
\varepsilon_{\phi\phi}=\frac{r-R}{R}
=\frac{1}{E}(\sigma_{\phi\phi}-\nu\sigma_{rr}-\nu\sigma_{zz}),
\label{Ec9}
\end{equation}

\noindent
where $E$ and $\nu$ are the Young coefficient and Poisson ratio of the
material the shell is made of. Substituting

\begin{equation}
\sigma_{zz}=-\frac{P}{2\pi re}\cos\theta ,
\label{Ec10}
\end{equation}

\noindent
$\cos\theta =[1+(dr/dz)^2]^{-1/2}$ and

\begin{equation}
\sigma_{rr}=-\frac{F_\phi \Delta\phi}{r\Delta\phi\Delta z}
=-\sigma_{\phi\phi}\frac{e}{r}\sqrt{1+\bigg(\frac{dr}{dz}\bigg)^2}
\label{Ec11}
\end{equation}

\noindent 
in Eq.~(\ref{Ec9}), and solving for $\sigma_{\phi\phi}$, it is
obtained that

\begin{equation}
\sigma_{\phi\phi}=\frac{1}{1+\nu\dfrac{e}{r}
\sqrt{1+\bigg(\dfrac{dr}{dz}\bigg)^2}}\left(E\,\frac{r-R}{R}
-\frac{\nu P}{2\pi er\sqrt{1+\bigg(\dfrac{dr}{dz}\bigg)^2}}\right).
\label{Ec12}
\end{equation}

\noindent 
Combining Eqs.~(\ref{Ec12}) and (\ref{Ec8}) one finally obtains

\begin{equation}
\frac{d^2r}{dz^2}
=-\dfrac{\dfrac{2\pi eE}{RP}(r-R)\sqrt{1+\bigg(\dfrac{dr}{dz}\bigg)^2}
-\dfrac{\nu}{r}}{1+\nu\dfrac{e}{r}
\sqrt{1+\bigg(\dfrac{dr}{dz}\bigg)^2}}\, .
\label{Ec13}
\end{equation}

The differential equation (\ref{Ec13}) gives the profile $r(z)$ of the
cylindrical shell deformed by the axial load $P$ when the proper
physical conditions are imposed to the solutions. The most evident
conclusion one can infer from Eq.~(\ref{Ec13}) is that the load $P$
cannot be null. Hence it corresponds to a deformation regime occurring
for high enough loads.

The unloaded cylindrical shell has a length $L$ in the axial
direction, which reduces to a smaller length $l$ by the applied load
$P$. The length reduction originates from the elastic compressive
strain of the deformed mantle and by the shape deformation itself
projected in the $z$--direction. If the elastic strain can be
neglected when compared with the effect of the shell deformation, the
length of the curved path adopted by the deformed shell will conserve
its original value $L$. Hence the physical solutions of
Eq.~(\ref{Ec13}) must satisfy

\begin{equation}
L=\int_{-l /2}^{l /2}dz\,\sqrt{1+\left(\frac{dr}{dz}\right)^2}.
\label{Ec14}
\end{equation}

\subsection{Present theoretical approach versus theory of elasticity}
\label{approach}

The equation for the shell profile was derived in the preceding
subsections from analysing the equilibrium of the forces exerted on a
representative elementary sector of the material medium, introducing
from the beginning the symmetry constraints. For our present purposes,
this procedure is much simpler and practical than the more standard
approach of readily introducing the system boundary conditions into
the general formalism of the theory of elasticity.  This is because
the latter approach oblige us to face complex problems we are not
interested in. In the theory of elasticity the free surfaces of the
shell are defined as surfaces where the stresses vanish, and the
theory furnishes the equations for calculating the detailed structure
of the stresses occurring in between. A recent paper by Zozulya and
Zhang \cite{Zozulya} provides a good example of this kind of precise
calculation of the deformation of cylindrical shells, together with a
detailed account of the stress fields inside the material. The cost
paid for such a complete solution is the introduction of numerical
methods from the very start.

However, it is not necessary here to know about the precise profiles
of the outer and inner surfaces of the shell, or how vary the stresses
inside. The forces $F_c(z)$ and $F_\phi (z)$ in Fig.~\ref{Fig3}
account for the integrated effect of these stresses, and their
equilibrium condition proves to be enough for determining the equation
for the mean shell profile, which has the advantage of being
closed--form.

\section{The profile equation in the buckling regime of deformation
and its solution}
\label{buckling}

The trivial solution $r=\text{constant}$ reduces Eq.~(\ref{Ec13}) to

\begin{equation}
E\frac{r-R}{R}=\nu\frac{P}{2\pi re} \qquad (r=\text{constant}),
\label{Ec15}
\end{equation}

\noindent
which can be interpreted as the Poisson effect on the shell circular
perimeter $2\pi r$ accompanying the homogeneous axial strain produced
by the applied compressive stress $P/(2\pi re)$. This uniform
solution, which is expected to be stable up to a critical load $P_B$,
preserves the cylindrical shape and produces only small geometrical
variations, even for big loads, because $E$ is usually very large 
(0.01--500 GPa). The buckled non trivial solutions of Eq.~(\ref{Ec13})
involve much larger deformations. In the post--buckling regime $r$
assumes values in the interval $R<r\le R+L/2$ when $l$ varies from
$l=L$ to $l=0$. The second term $\nu /r$ in the numerator of the right
hand side of Eq.~(\ref{Ec13}) is comparable with the first one only
when $|r-R|/R\ll 1$. As long as $r$ departs from $R$ beyond the range
of the purely elastic distortions, the term $\nu /r$ becomes
negligibly small when compared with the one proportional to $E$. Hence
it is advisable to distinguish between elastic strains and geometric
changes and write

\begin{equation}
\frac{d^2r}{dz^2}
=-\frac{2\pi eE}{PR}(r-R)\sqrt{1+\left(\frac{dr}{dz}\right)^2}
\quad\text{(buckling regime)}. 
\label{Ec16}
\end{equation}

\noindent
Ec.~(\ref{Ec16}) also assumes a thin enough shell to neglect the
second term in the denominator of the right hand side of
Ec.~(\ref{Ec13}).  Although this non-linear equation is not listed in
the specialized treatises on elliptic functions and integrals
\cite{Gradshteyn,Byrd}, it will be shown next that its exact solution
is a Jacobi elliptic sine function, which holds for any load and
deformation state, including the limit in which the shell has been
completely flattened by the applied force.

Defining

\begin{equation}
y(z)=r(z)-R, \quad p=\frac{2\pi eE}{PR}, \quad R=\text{constant},
\label{Ec17}
\end{equation}

\noindent
Eq.~(\ref{Ec16}) reads

\begin{equation}
y''=-py\sqrt{1+y'^2},
\label{Ec18}
\end{equation}

\noindent
and the substitution 

\begin{equation}
y(z)=\frac{1}{\sqrt{p}}\, u(\sqrt{p}\, z)
\label{Ec19}
\end{equation}

\noindent
turns it into

\begin{equation}
u''=-u\sqrt{1+u'^2}.
\label{Ec20}
\end{equation}

\noindent
Multiplying both sides of this equation by $u'$ yields the integrable
form

\begin{equation}
\frac{u''u'}{\sqrt{1+u'^2}}=-uu',
\label{Ec21}
\end{equation}

\noindent
which can be solved to give

\begin{equation}
\sqrt{1+u'^2}=-\frac{1}{2}u^2+\frac{1}{2}u_0^2+1,
\label{Ec22}
\end{equation}

\noindent
where $u_0=u(0)$. Because of the symmetry with respect to the origin,
$u(\sqrt{p}\, z)$ must have its maximum at $z=0$ and the integration
constant was chosen so that $u'(0)=0$. Denoting now

\begin{equation}
\zeta =\sqrt{p}\, z, \quad v(\zeta)=\frac{u(\zeta)}{u_0},
\quad k^2=\frac{(u_0/2)^2}{1+(u_0/2)^2}\, ,
\label{Ec23}
\end{equation}

\noindent
Eq.~(\ref{Ec22}) can be rewritten as

\begin{equation}
\frac{dv}{d\zeta}=
\sqrt{(1-v^2)\bigg(1+\frac{u_0^2}{4}-\frac{u_0^2}{4}\, v^2\bigg)}\, ,
\label{Ec24}
\end{equation}

\noindent
or

\begin{equation}
\sqrt{1-k^2}\,\frac{dv}{d\zeta}=\sqrt{(1-v^2)(1-k^2v^2)}\, .
\label{Ec25}
\end{equation}

\noindent
Inverting this equation and integrating with respect to $v$ one has

\begin{equation}
\frac{1}{\sqrt{1-k^2}}(\zeta +\zeta_1)
=\int_{\displaystyle\, 0}^{\displaystyle\, v}
\frac{dv}{\sqrt{(1-v^2)(1-k^2v^2)}}=F(v,k),
\label{Ec26}
\end{equation}

\noindent
where $F(v,k)$ is the incomplete elliptic integral of the first kind
with modulus $k$, ($0\le k\le 1$) \cite{Gradshteyn,Byrd}, and
$\zeta_1$ is an integration constant. The same symbol was used for the
integration variable and the upper integration limit to simplify the
notation.

The inverse function of the incomplete elliptic integral $F(v,k)$ is
known as the Jacobi elliptic sine function sn. Hence Eq.~(\ref{Ec26})
is equivalent to

\begin{equation}
v=\text{sn }\bigg(\frac{\zeta +\zeta_1}{\sqrt{1-k^2}}\, ,k\bigg)
=\text{sn }\bigg(\sqrt{\frac{p}{1-k^2}}\, (z+z_1),k\bigg).
\label{Ec27}
\end{equation}

\noindent
Function $\text{sn }(x,k)$ takes values in the interval $[-1,1]$ and
is periodic in $x$ with period $4K(k)$, where $K(k)=F(1,k)$ is the
complete elliptic integral of the first kind. Also, $\text{sn }(0,k)=
\text{sn }(2K,k)=0$ and $\text{sn }(K,k)=1$. The Jacobi sine function
is symmetric with respect to $x=K$ and has a maximum there.

Therefore, the solution satisfying the boundary conditions
$y(\pm l/2)=0$ is such that

\begin{equation}
\sqrt{\frac{p}{1-k^2}}=\frac{2K(k)}{l}
\quad\text{and}\quad z_1=\frac{l}{2}.
\label{Ec28}
\end{equation}

\noindent
Taking $\sqrt{p}$ and $z_1$ from these equations, and recalling the
third of Eqs.~(\ref{Ec23}), which gives $u_0/2=k/\sqrt{1-k^2}$, one
has that the profile of the buckled cylindrical shell is given by

\begin{equation}
y(z)=r(z)-R=\frac{k}{(1-k^2)K(k)}\,\text{sn }\bigg[\frac{2K(k)}{l}
\bigg(z+\frac{l}{2}\bigg),k\bigg].
\label{Ec29}
\end{equation}

Combining the first Eq.~(\ref{Ec28}) with the definition (\ref{Ec17})
of the constant $p$, a relation between the applied force $P$ and $k$
follows

\begin{equation}
P=\frac{\pi eEl^2}{2R(1-k^2)K^2(k)}.
\label{Ec30}
\end{equation}

\noindent
Therefore, it rests just to determine the meaning of $k$ to have a
complete solution of our problem. In the next subsection it will be
shown that $k^2$ is essentially the relative axial deformation
$\varepsilon =(L-l)/L$ of the cylindrical shell. The elliptic integral
$K(k)$ can be calculated quite easily from the defining integral or
the series

\begin{equation}
K(k)=\frac{\pi}{2}\bigg[1
+\sum_{n=1}^\infty\bigg(\frac{(2n-1)!!}{2^nn!}\bigg)^2k^{2n}\bigg].
\label{Ec31}
\end{equation}

\begin{figure}[h!]
\begin{center}
\includegraphics[width=6cm]{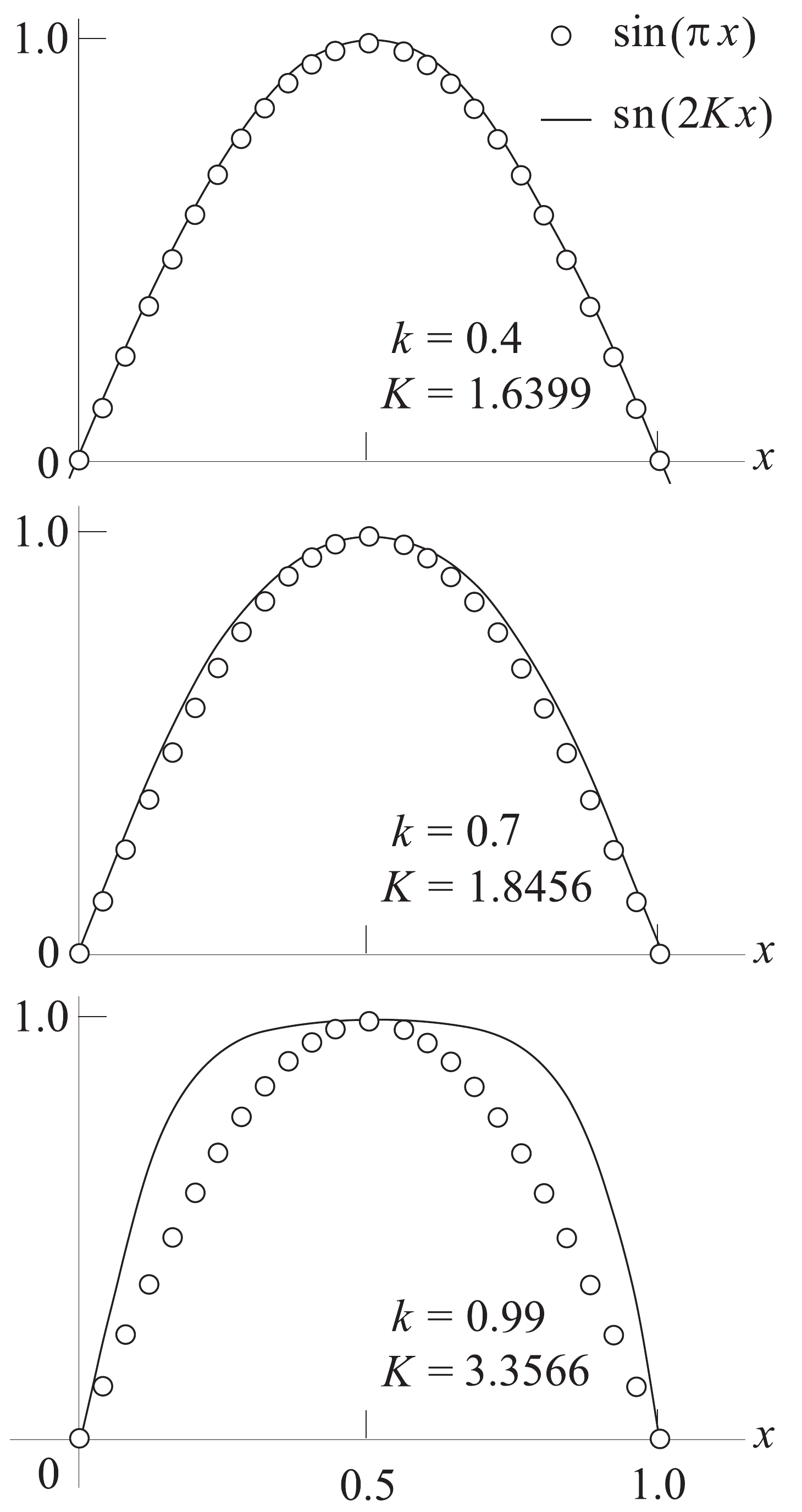}
\caption{\label{Fig7} The Jacobi elliptic sine function (solid lines)
for three values of the modulus $k$, compared with the trigonometric
sine function (circles). For $k<0.7$ the two functions are very
similar. At $k>0.7$ the elliptic function is sensibly broader and goes
to a square wave in the limit $k=1$. Physically, $k=1$ corresponds to
the situation in which the cylindrical mantle has been completely
flattened by the applied force $P$ and $l=0$.}
\end{center}
\end{figure}

Fig.~\ref{Fig7} shows the Jacobi sine function $\text{sn }(2Kx,k)$ for
three values of $k$. The graphs makes apparent that the Jacobi
elliptic function goes from $\sin (\pi x)$ to a square wave when the
modulus $k$ goes from 0 to 1. The latter situation corresponds to the
final collapse of the cylindrical mantle, when $l=0$.

\subsection{The equation for the modulus $k$}\label{modulus}

When $R=\text{constant}$ one has $r'(z)=y'(z)=u'(\sqrt{p}\, z)$, hence
$(1+r'^2)^{1/2}$ can be replaced by $(1+u'^2)^{1/2}$ in the general
condition expressed by Eq.~(\ref{Ec14}). Making this and then
substituting Eq.~(\ref{Ec22}),

\begin{equation}
L=\int_{-l/2}^{l/2}\sqrt{1+u'^2}\, dz=-\frac{1}{2}\int_{-l/2}^{l/2}
u^2(\sqrt{p}\, z)\, dz+\frac{1}{2}u_0^2l+l .
\label{Ec32}
\end{equation}

\noindent
Recalling now
 
\begin{equation}
u(\sqrt{p}\, z)=
u_0\,\text{sn }\bigg[\frac{2K}{l}\bigg(z+\frac{l}{2}\bigg),k\bigg],
\quad u_0=\frac{2k}{\sqrt{1+k^2}},
\label{Ec33}
\end{equation}

\noindent
Eq.~(\ref{Ec32}) can be written as

\begin{equation}
\frac{L-l}{l}=\frac{2k^2}{1-k^2}\left( 1-\frac{1}{2K(k)}\int_0^{2K(k)}
\text{ sn}^2\, (\zeta ,k)\, d\zeta\right).
\label{Ec34}
\end{equation}

\noindent
The integral appearing in this equation has been solved in terms of
the elliptic functions and one has the mathematical identity
\cite{Gradshteyn}

\begin{equation}
\int\,\text{sn}^2\,\zeta\, d\zeta
=\frac{1}{k^2}[u-E(\text{am }\zeta,k)],
\label{Ec35}
\end{equation}

\noindent
where $E(\phi ,k)$, $0\le\phi\le\pi/2$, is the second kind incomplete
elliptic integral with modulus $k$. In the usual notation of the
theory of elliptic integrals the amplitude am means $\text{am }\zeta=
\arcsin(\text{sn }\zeta)$. Care must be taken in replacing properly
the integration limits in the indefinite integral (\ref{Ec35}) because
$0\le\text{am }\zeta\le\pi/2$, and $\text{am }K=\pi/2$. Therefore,
$2K$ is outside the domain in which the amplitude function is defined.
To overcome this difficulty one can take advantage of the symmetry of
$\text{sn }(x,k)$ with respect to $x=K$ and write

\begin{equation}
\int_0^{2K}\,\text{sn}^2\,\zeta\, d\zeta
=2\int_0^{K}\,\text{sn}^2\,\zeta\, d\zeta
=\frac{2}{k^2}[K-E(\pi/2,k)]
\label{Ec36}
\end{equation}

\noindent
replacing $\text{am }K=\pi/2$. $E(\pi/2,k)=E(k)$ is the complete
elliptic integral of the second kind, modulus $k$, defined by the
integral \cite{Gradshteyn}

\begin{equation}
E(k)=\int_0^{\pi/2}\sqrt{1-k^2\sin^2\phi}\, d\phi,
\label{Ec37}
\end{equation}

\noindent
or the series

\begin{equation}
E(k)=\frac{\pi}{2}\bigg[ 1-\sum_{n=1}^\infty
\bigg(\frac{(2n-1)!!}{2^nn!}\bigg)^2\frac{k^{2n}}{2n+1}\bigg].
\label{Ec38}
\end{equation}

\noindent
The term $D(k)=(1/k^2)[K(k)-E(k)]$ can be evaluated either by
combining the power series (\ref{Ec31}) and (\ref{Ec38}) as

\begin{equation}
D(k)=\frac{K(k)-E(k)}{k^2}
=\frac{\pi}{2}\sum_{n=0}^\infty\frac{2(n+1)}{2n+1}
\bigg(\frac{(2n+1)!!}{2^{n+1}(n+1)!}\bigg)^2k^{2n},
\label{Ec39}
\end{equation}

\noindent 
or solving the integral

\begin{equation}
D(k)=\int_0^{\pi/2}\frac{\sin^2\phi\, d\phi}{\sqrt{1-k^2\sin^2\phi}}\, ,
\label{Ec40}
\end{equation}

\noindent
which follows directly from the definitions of $K(k)$ and $E(k)$.

\begin{figure}[h!]
\begin{center}
\includegraphics[width=7cm]{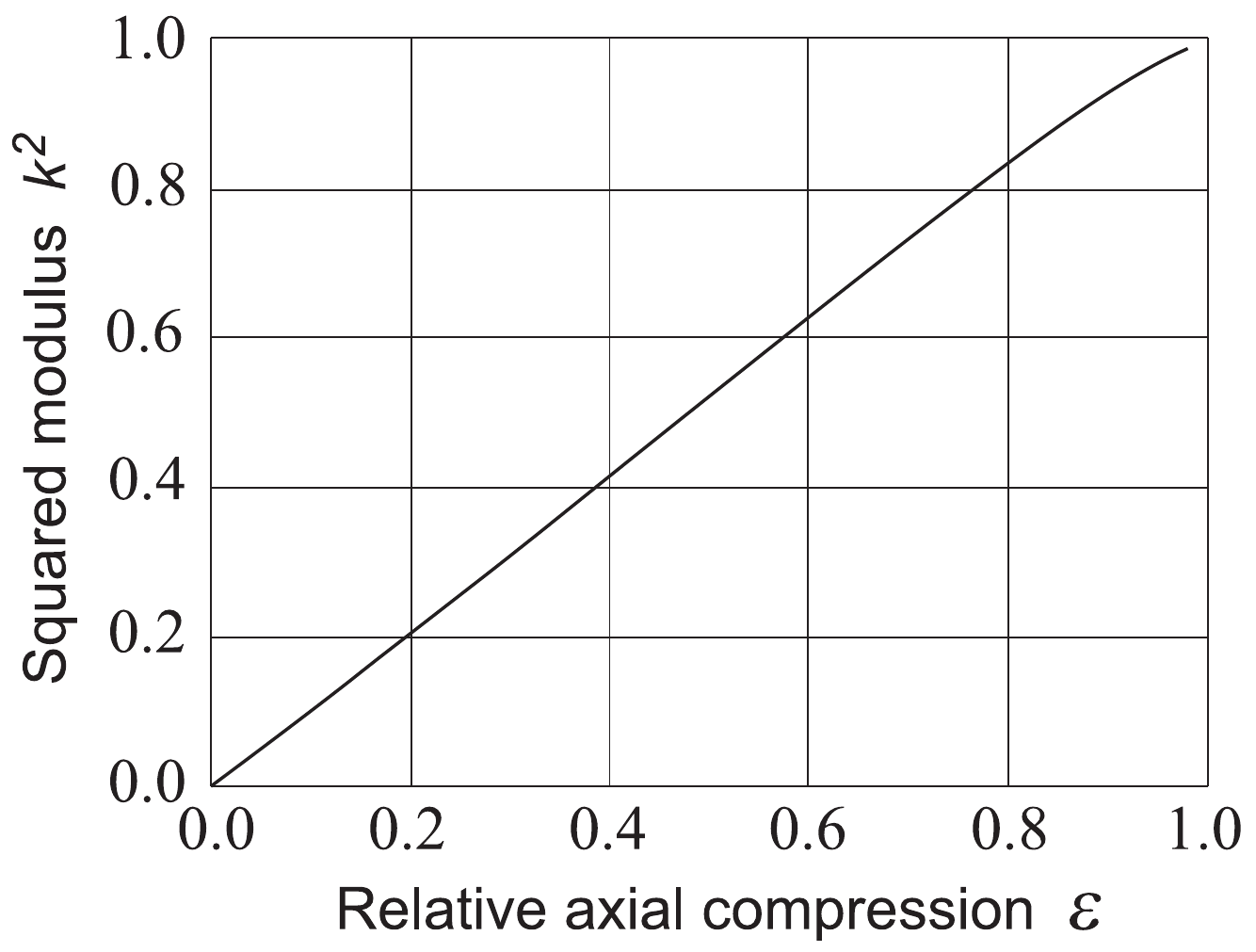}
\caption{\label{Fig8} The squared modulus $k^2$ and the relative axial
compression $\varepsilon =(L-l)/L$, as given by Eq.~(\ref{Ec41}).}
\end{center}
\end{figure}

Eq.~(\ref{Ec34}) then becomes

\begin{equation}
\frac{\varepsilon}{1-\varepsilon}=
\frac{2k^2}{1-k^2}\left( 1-\frac{D(k)}{K(k)}\right),
\quad \varepsilon =\frac{L-l}{L},
\label{Ec41}
\end{equation}

\noindent
which determines exactly the modulus $k$ as a function of only the
relative axial compression $\varepsilon =(L-l)/L$. The function
$k=f(\varepsilon)$ does not depend on the characteristics of the
cylindrical shell and is shown in Fig.~\ref{Fig8}. The amazingly
simple approximate rule 

\begin{equation}
k^2\approx\varepsilon
\label{Ec42}
\end{equation}

\noindent
holds over the whole range of $\varepsilon$ with a 5\% maximum error.

\section{Critical load and reaction force}
\label{critical}

The deformation regime studied in the preceding sections takes place
for a constant non--vanishing axial load $P$, given by
Eq.~(\ref{Ec30}). This is because it was assumed from the very
beginning that loading curves the cylindrical shell profile and
Eqs.~(\ref{Ec29}), (\ref{Ec30}) and (\ref{Ec41}) hold at buckling and
in the post--buckling regime of deformation. The variable
$\varepsilon$ measures only the deformation induced relative
displacement of the edges along the axial direction, and the much
smaller elastic distortions have been neglected. The force

\begin{equation}
P=\frac{\pi eEL^2(1-\varepsilon)^2}{2R(1-k^2)K^2(k)}
\label{Ec43}
\end{equation}

\noindent
has an absolute maximum $P_B$ at $\varepsilon =k=0$. Recalling
$K(0)=\pi/2$, the maximal reaction force reads

\begin{equation}
P_B=\frac{2eEL^2}{\pi R},
\label{Ec44}
\end{equation}

\noindent
and one can write 

\begin{equation}
\frac{P(\varepsilon)}{P_B}
=\frac{\pi^2}{4}\frac{(1-\varepsilon)^2}{(1-k^2)K^2(k)}.
\label{Ec45}
\end{equation}

\begin{figure}[h!]
\begin{center}
\includegraphics[width=7cm]{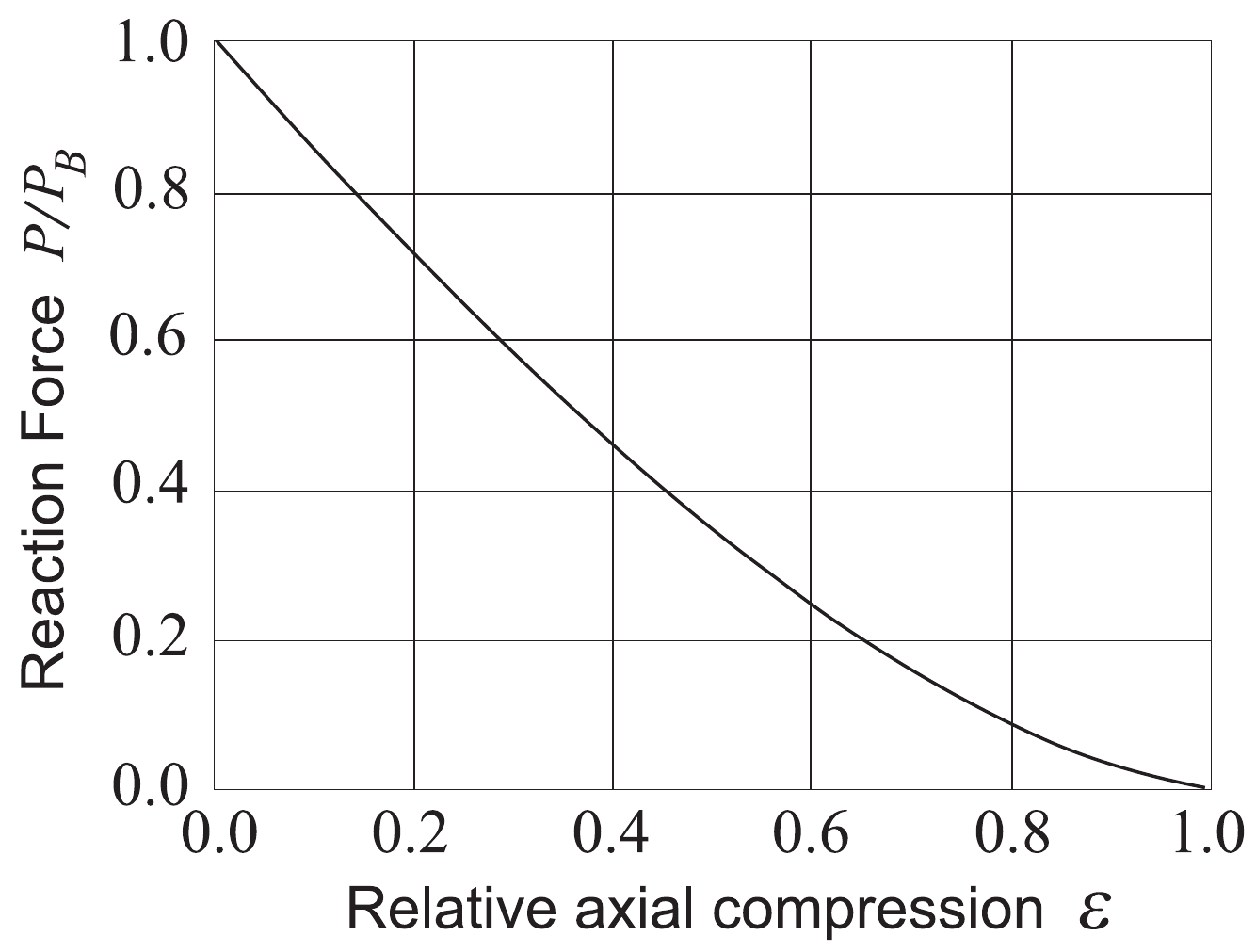}
\caption{\label{Fig9} Reaction force $P$ relative to the maximum $P_B$
as a function of axial compression $\varepsilon =(L-l)/L$ , as given
by Eq.~(\ref{Ec45}).}
\end{center}
\end{figure}
 
\begin{figure}[h!]
\begin{center}
\includegraphics[width=7cm]{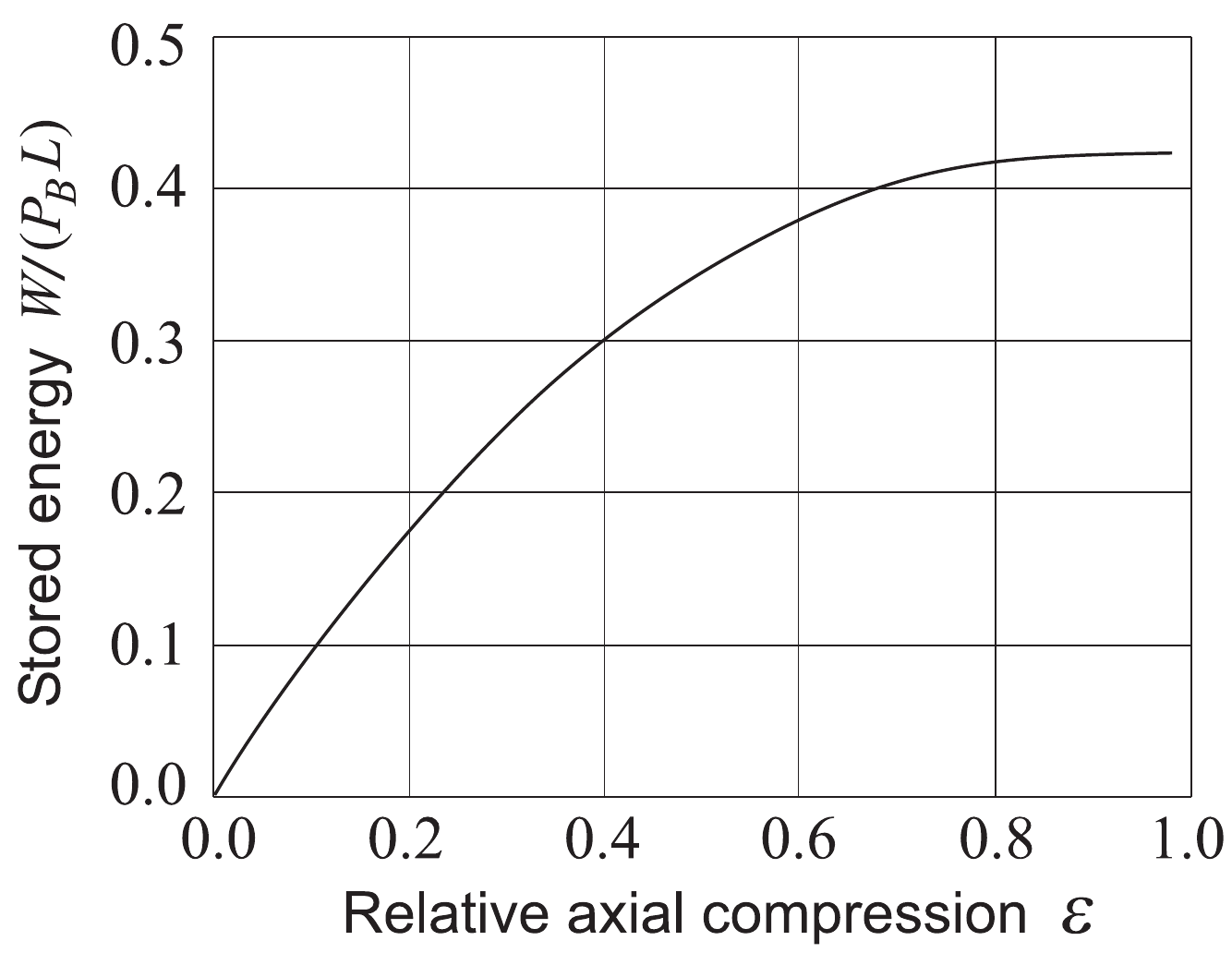}
\caption{\label{Fig10} Energy absorbed by the cylindrical shell in the
post--buckling deformation regime, in units of $P_BL$ and for any
relative axial compression $\varepsilon$.}
\end{center}
\end{figure}

\noindent
Fig.~\ref{Fig9} shows $P/P_B$, which according to Eqs.~(\ref{Ec45})
and (\ref{Ec41}) is an universal function of $\varepsilon$.
Fig.~\ref{Fig10} displays the energy stored by the shell in the
post--buckling deformation regime, obtained integrating
Eq.~(\ref{Ec45}) with respect to $\varepsilon$.

\section{Homogeneous and buckled deformation regimes}

The problem we are interested in is a deformation in which the
cylindrical shell is gradually compressed along its main symmetry axis.
In a first stage the compression is elastic and conserves the
cylindrical shape. The relative displacement $x$ of the edges is given
by Hookes's law

\begin{equation}
\frac{P}{2\pi Re}=E\frac{x}{L_0},
\label{Ec46}
\end{equation}

\noindent
where $L_0$ is the original length of the unstrained cylinder. Once
the elastic distortion $x$ takes a critical value $x_B$, such that the
applied force $P$ reaches the threshold strength $P_B$, the cylinder
buckles and Eq.~(\ref{Ec43}) starts holding, instead of
Eq.~(\ref{Ec46}). At the critical deformation $x_B$ the surface just
starts to acquire the barrel shape, the two deformation regimes
coexist and Eqs.~(\ref{Ec43}) and (\ref{Ec46}) are both valid for
$L=L_0-x_B$ and $\varepsilon =0$ (notice that $L$ stands for the the
shell length when buckling is just initiated). Thus the axial
deformation $x_B$ at which the buckling regime sets in can be obtained
from combining Eqs.~(\ref{Ec44}) and (\ref{Ec46}) to eliminate $P_B$.
This yields

\begin{equation}
\frac{x_B}{L_0}=
1+\left(\frac{\pi R}{\sqrt{2}L_0}\right)^2
-\frac{\pi R}{\sqrt{2}L_0}
\sqrt{\left(\frac{\pi R}{\sqrt{2}L_0}\right)^2+2},
\label{Ec47}
\end{equation}

\noindent
which is the equation for the critical elastic strain $x_B/L_0$. The
critical load is

\begin{equation}
P_B=2\pi ReE\,\bigg[ 1+\left(\frac{\pi R}{\sqrt{2}L_0}\right)^2
-\frac{\pi R}{\sqrt{2}L_0}
\sqrt{\left(\frac{\pi R}{\sqrt{2}L_0}\right)^2+2}\bigg].
\label{Ec48}
\end{equation}

\section{Conclusions relative to the marine application}
\label{example}

To illustrate why vehicle old tires are functional as dock dampers in
the light of the preceding equations, consider a 27''$\times$ 47''
tire, originally designed for heavy mining trucks. The dimensions of
the tread are $L_0=0.75\,\text{m}$, $R=1.25\,\text{m}$ and
$e=0.10\,\text{m}$. Young's effective modulus for tire rubber with
added reinforcements can be estimated as $E=10\text{ MPa}$.
The purpose of this calculation is fundamentally illustrative, to
acquire a feeling of the magnitudes involved. Replacing these
constants in Eqs.~(\ref{Ec47}) and (\ref{Ec48}) one has that

\begin{equation}
x_B=2.55\,\text{cm}\quad P_B=267\,\text{kN}
=27.3\,\text{ton}.
\label{Ec49}
\end{equation}

\begin{figure}[h!]
\begin{center}
\includegraphics[width=7.5cm]{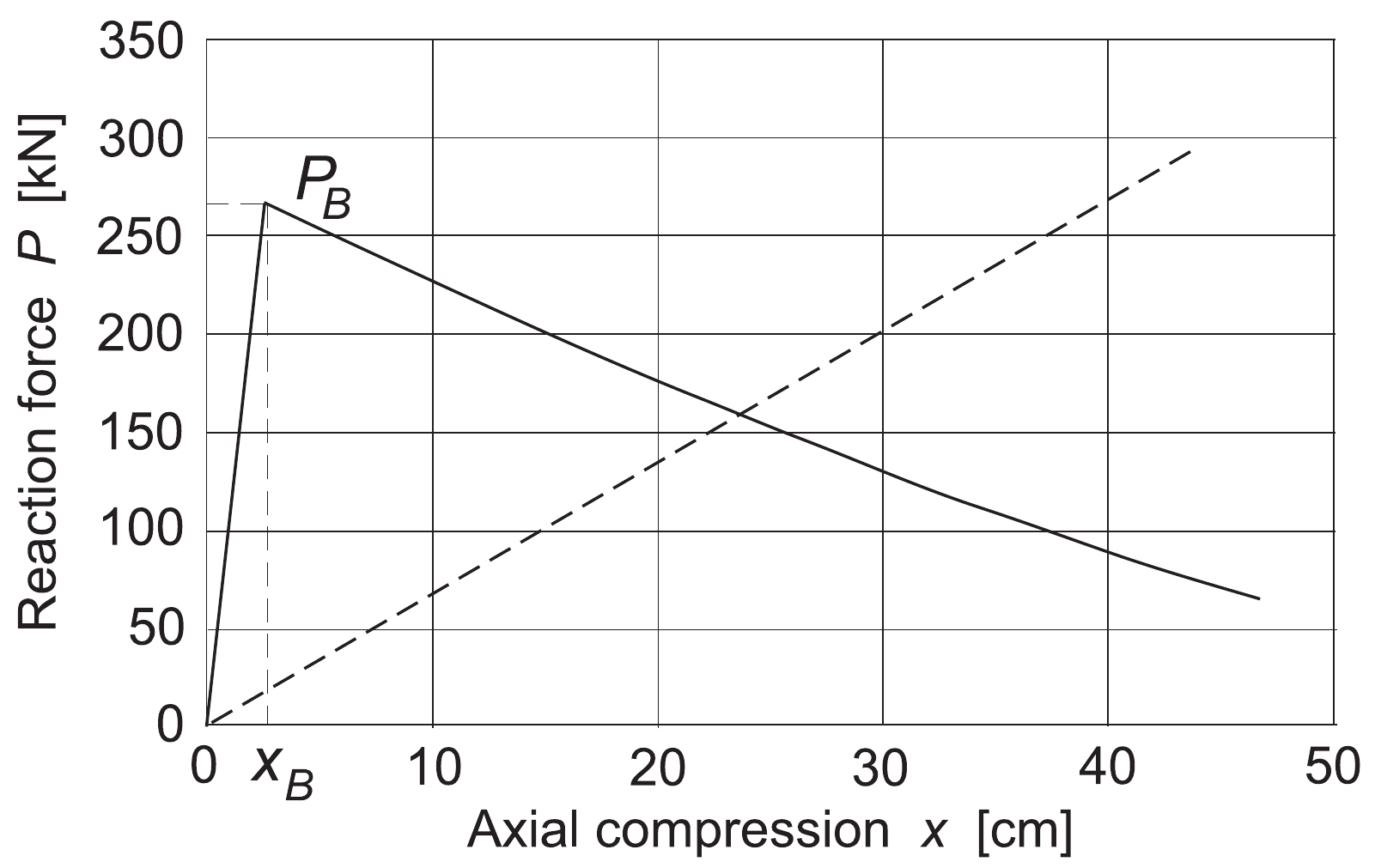}
\caption{\label{Fig11} Axial reaction force $P$ exerted by a tire with
tread dimensions $L_0=0.75\,\text{m}$, $R=1.25\,\text{m}$ and
$e=0.10\,\text{m}$, and effective Young's modulus $E=10\,\text{MPa}$,
as a function of the axial compression $x$. The broken line shows the
behaviour of a spring or any other strictly elastic medium chosen to
yield a maximum force comparable with the tire.}
\end{center}
\end{figure}

Thus, when the tire is progresively compressed, the reaction force
grows linearly with strain up to the critical strength
$P_B=267\,\text{kN}$ at $x_B=2.55\,\text{cm}$, with the tread
conserving its cylindrical shape. As compression continues, the tread
curves and the radius turns dependent on $z$ according to
Eq.~(\ref{Ec29}). Fig.~\ref{Fig11} shows the behaviour of the reaction
$P$ of the strained tire in the two regimes, the linear and the
buckled one (solid line). In the latter regime force $P$ obeys
Eq.~(\ref{Ec45}) with $L=L_0-x_B$. The broken line is the force
exerted by a strictly elastic system, like a spring matrix, introduced
here for the sake of comparison.

\begin{figure}[h!]
\begin{center}
\includegraphics[width=7.5cm]{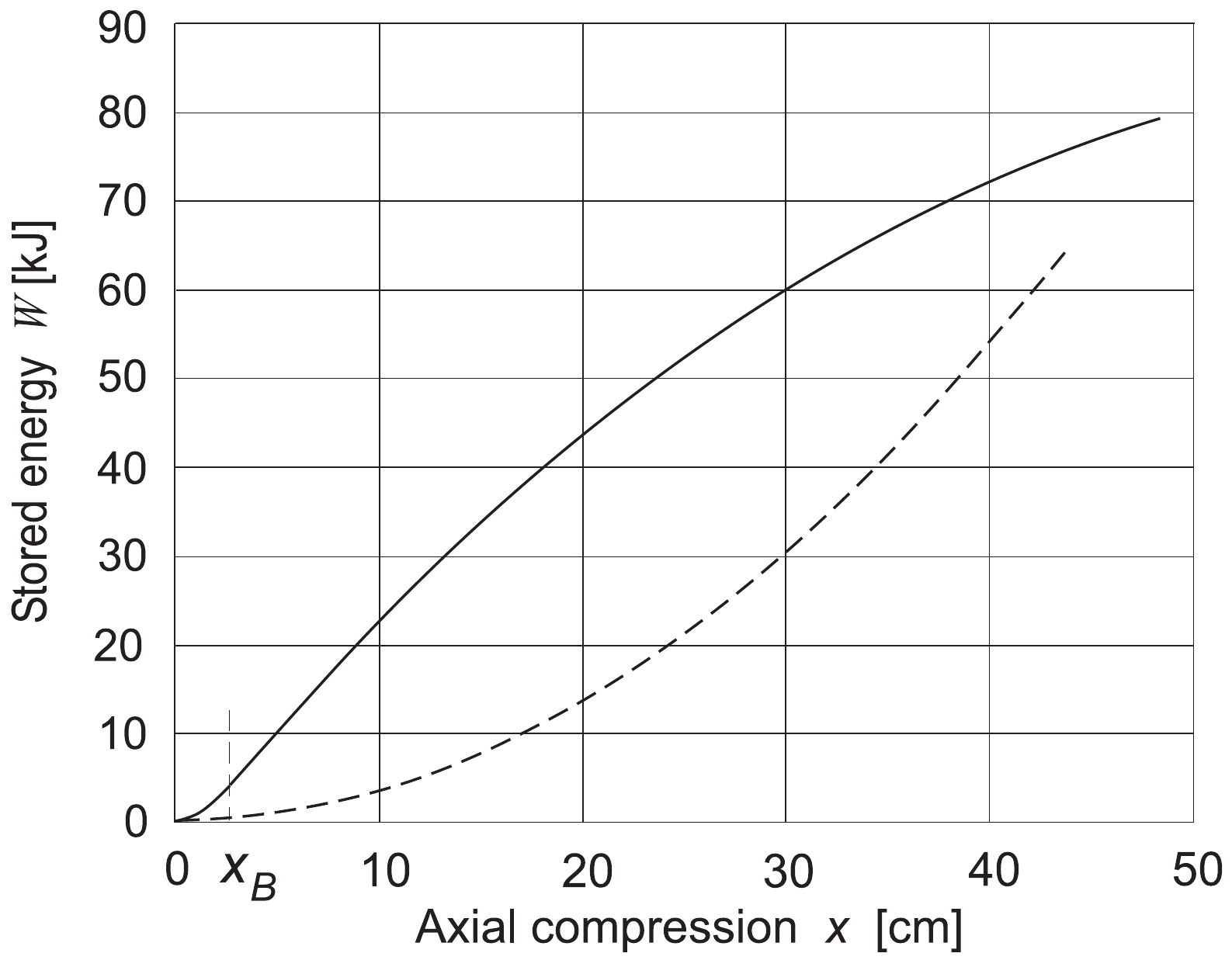}
\caption{\label{Fig12} The solid curve represents the energy absorbed
by the tire of Fig.~\ref{Fig11} when compressed in a distance $x$. The
broken curve is the same magnitude for the strictly elastic system.
The smaller capacity of the latter for compressions in the range
between 0 and 40 cm is apparent.}
\end{center}
\end{figure}

The stopping power of the tire when used as a dock bumper device is
given by the area $W(x)$ under the curve of the reaction force. The
solid line in Fig.~\ref{Fig12} shows the curve of the energy stored
$W(x)$ by the tire as a function of the compressive displacement $x$.
The broken line displays the same magnitude for the purely elastic
shock energy absorber.

Figs.~\ref{Fig11} and \ref{Fig12} show the features that make tires
good dock bumpers:\\
(a) The reaction force has a well defined maximum $P_B$, which ensures
that the energy absorption device will not damage the colliding
structures.\\
(b) Once the maximum reaction is reached, the reaction $P$ decreases
smoothly with the displacement $x$, and energy storage $W(x)$ keeps
high. This is the physical magnitude determining the stopping power
of the device.\\
(c) Function $W(x)$ increases monotonically and rapidly with $x$,
particularly for small values of $x$. This means that the stopping
action of the axially compressed tire rapidly increases and keeps
high at any strain, even for small ones. The superiority of the
cylindrical shell (solid line) over a strictly elastic system (broken
line) is shown quite dramatically in Fig.~\ref{Fig12}. The latter
reaches an acceptable stopping ability at the cost of increasing the
reaction force to a dangerous level.\\
(d) As $x$ takes relatively large values, the reaction force $P$
of the cylindrical bumper decreases to a fraction of its initial
strength, reducing this way its ability to produce rebound. The
elastic bumper produces the opposite effect.    

Think for instance of a ship of mass $m=55,500\,\text{ton}$ being
pushed by the tugs to approach normally the dock at a speed
$v=0.24\,\text{knot}=0.1234\,\text{m/s}$ (a rather high speed). The
kinetic energy of the ship is then $423\,\text{kJ}$. Assuming a safe
tire compression of 0.4 m, the graph in Fig.~\ref{Fig12} shows that
each tire can store $W(0.40\,\text{m})=72.5\,\text{kJ}$, and then a
minimum of six units would be necessary for safely docking the vessel.
On the other hand, the ship side is specified to support a maximal
load of $147\,\text{kN/m}^2$. The peak pressure the tires can exert on
the ship side is $P_B/(\pi R^2)=54.4\,\text{kN/m}^2$, which is close
to 1/3 the specified maximum.

\end{document}